\documentclass[12pt]{article}
\usepackage{eurosym}
\usepackage[left=2.2cm, right=2.2cm, top=2.2cm, bottom=2.2cm]{geometry}
\usepackage{amsmath,amsfonts,amssymb,amsthm}
\usepackage{lscape}
\usepackage{graphicx}
\usepackage{natbib}
\usepackage{amsmath,amssymb,graphicx,setspace,rotating}
\usepackage[outdir=./]{epsfig} 
\usepackage{pdflscape}
\usepackage[section]{placeins}   
\usepackage{color}  
\usepackage{threeparttable} 
\usepackage{bm} 
\usepackage{algorithm}
\usepackage[noend]{algpseudocode}
                  
\DeclareMathOperator*{\plim}{plim}

\newtheorem{definition}{Definition}

\makeatletter
\newcommand*\bigcdot{\mathpalette\bigcdot@{.5}}
\newcommand*\bigcdot@[2]{\mathbin{\vcenter{\hbox{\scalebox{#2}{$\m@th#1\bullet$}}}}}
\makeatother

\newtheorem*{example}{Example}
\newtheorem{example*}{Example}

\def\1{1\!{\rm l}}

\def\y{\mathbf{y}}
\def \zb{\mathbf{z}}

\newcommand{\dt}{\text{d}}

\newcommand{\z}{\mathbf{z}}
\newcommand{\dx}{\text{d}}

\begin{document}

\title{Robust Approximate Bayesian Computation: An Adjustment Approach}
  \vspace{0.7cm}

\date{\today}

\author{ David T. Frazier\footnote{ 
		Department of Econometrics and Business Statistics and the Australian Centre of Excellence for Mathematical and Statistical Frontiers, Monash University    
		(\texttt{david.frazier@monash.edu}).}, Christopher Drovandi\footnote{ School of Mathematical Sciences and the Australian Centre of Excellence for Mathematical and Statistical Frontiers, Queensland University of Technology (\texttt{c.drovandi@qut.edu.au})} and Ruben Loaiza-Maya\footnote{ 
		Department of Econometrics and Business Statistics and the Australian Centre of Excellence for Mathematical and Statistical Frontiers, Monash University    
		(\texttt{ruben.LoaizaMaya@monash.edu}).}}

\maketitle

\begin{abstract}
We propose a novel  approach to approximate Bayesian computation (ABC) that seeks to cater for possible misspecification of the assumed model. This new approach can be equally applied to rejection-based ABC and to popular regression adjustment ABC. We demonstrate that this new approach mitigates the poor performance of regression adjusted ABC that can eventuate when the model is misspecified. In addition, this new adjustment approach allows us to detect which features of the observed data can not be reliably reproduced by the assumed model. A series of simulated and empirical examples illustrate this new approach.   
\end{abstract}

\vspace{.5cm}


\vspace{.5cm}

\section{Introduction}

The now common use of complex models has led to the rise of approximate Bayesian methods, with the goal of these methods being to construct a useful {{approximation}} to the exact Bayesian posterior distribution. Application of these approximate methods has grown dramatically over the last decade, and they now occupy an important place in the armoury of the practising statistician. One of the most popular approximate Bayesian methods is the method of approximate Bayesian computation (ABC); for a handbook-style treatment on ABC see \cite{sisson2018}.

ABC eschews calculation of the likelihood in favour of simulation from the assumed model. In contrast to exact Bayesian methods, which must explicitly calculate the likelihood, or an unbiased estimator thereof, ABC only
requires that one is able to \textit{simulate} pseudo data sets from the assumed model; parameter values that generate data sets which are close to the observed data are retained and used to estimate the posterior distribution. 

Due to the curse of dimensionality inherent in ABC \citep{blum2010approximate},  inference is often conditioned on a low-dimensional summary statistic of the full data, to maintain reasonable computation times, and/or post-processing of the initial ABC output is employed. For example, it is common to adjust the original ABC output using a linear or nonlinear regression (see, e.g., \citealp{Beaumont2025} and \citealp{blumF2010non}). In cases of correct model specification regression adjustment ABC approaches can lead to more accurate posterior approximations (\citealp{LF2016b}).

While ABC is currently applied in many different research areas, \cite{frazier2017model} have questioned the blind application of ABC in settings where the assumed model may not be an accurate representation of the true data generating process (DGP); i.e., when the assumed model is misspecified. In particular, these authors give both theoretical and empirical evidence that when the assumed model is misspecified ABC-based inference may yield misleading conclusions. Moreover, such behavior can be exacerbated by the application of regression adjustment approaches. More specifically, when the model is misspecified regression adjustment ABC can yield posteriors with poor coverage and unstable point estimators.

Motivated by the behavior of ABC in misspecified models, we propose a novel adjustment approach to ABC that yields robust inferences in misspecified models. Through a sequence of examples, we demonstrate that when this new approach is applied in conjunction with regression adjustment ABC, the resulting procedure completely ameliorates the poor performance of regression adjustment ABC that is sometimes observed in misspecified models; delivering posteriors with well-behaved point estimators and good frequentest coverage (for a well-defined pseudo-true value). Lastly, we demonstrate that this new ABC approach can pinpoint which of the summary statistics used in the analysis are misspecified, in a sense that we make more precise later.

The first approach we propose adjusts the location of the summaries by adding a vector of parameters that ``soak up'' the model misspecification. An additional adjustment approach is considered that weights the individual summaries used in the analysis in such a way that if the simulated and observed summaries do not agree, the overall distance can still be made small.

This new adjustment approach to ABC is inspired by the adjustment idea in \cite{FD2019}. In the context of Bayesian synthetic likelihood (BSL, \citealp{wood2010}, \citealp{price2018bayesian}), \cite{FD2019} demonstrate that when the model is misspecified BSL can deliver misleading inference. To circumvent this issue, \cite{FD2019} augment the BSL posterior with additional parameters that ``soak up'' the model misspecification. While the overall idea behind the approach proposed herein is similar to \cite{FD2019}, the differences between the BSL and ABC posterior targets requires important differences between the two approaches. We forgo an in-depth  comparison between the two approaches until Section three.

The remainder of the paper is organized as follows. In Section two we give a brief overview of ABC and discuss the issue of model misspecification in ABC. Section three presents our robust approach to ABC, and demonstrates in a toy example that this approach delivers reliable performance under model misspecification. Section four contains a mix of Monte Carlo and empirical results that further demonstrate the performance of this robust ABC approach. Section five concludes.

\section{Approximate Bayesian Computation and Model Compatibility}
\subsection{Approximate Bayesian Computation Framework}

The modeler observes data $\mathbf{y}:=(y_{1},...,y_{n})^{\prime }$ and wishes to conduct Bayesian inference on a complex class of parametric models $\{\theta\in\Theta:P_{\theta}\}$, where $\Theta\subset\mathbb{R}^{p}$ represents the parameter space for the unknown parameter $\theta$, and where $P_\theta$ denotes the probability measure of the model, and $p_\theta$ its density function. Our prior beliefs over $\theta$ are represented by the density $\pi(\theta)$. From the observed data $\y$, the model $P_\theta$, and our prior belief $\pi(\theta)$, Bayes Theorem delivers the cornerstone of Bayesian statistics: the posterior density $$\pi(\mathbf{\theta |y})\propto p_\theta(\mathbf{y})\pi(\mathbf{	\theta }).$$ 

Generally speaking, exact Bayesian inference (up to Monte Carlo error) requires that  $\pi(\mathbf{\theta |y})$ be available
in closed-form (i.e., analytically), at least up to the constant of
proportionality. On the other hand, {\textit{approximate}} Bayesian inference schemes generally remain applicable  in cases where $P_{\theta}$ (or $\pi(\theta )$, or both)
cannot be expressed in an analytic form, or are computationally too costly to employ in more standard  algorithms, such as Markov chain Monte Carlo.

The aim of ABC is to build a reliable approximation to $\pi(\mathbf{\theta |y})$ in cases where $P_\theta$ is not accessible. ABC is predicated on the belief that the observed data $\mathbf{y}$
is drawn from one of the constituent members in the class $\{{\theta \in
	\Theta }:P_{{\theta }}\}$, and conducts inference on the unknown $\theta$ by first drawing $\theta\sim\pi(\theta)$, then simulating pseudo-data $\mathbf{z}$, $\mathbf{z}:=(z_{1},...,z_{n})^{'}\sim P_{\theta}$,  and ``comparing'' $\mathbf{z}$ with the observed data $\mathbf{y}$. In most cases, this comparison is carried out using a vector of summary statistics $\eta(\cdot)$ and a metric $d\{\cdot,\cdot\}$. Simulated values of $\theta$ are then accepted, and used to build an approximation to the exact posterior, if the distance $d\{\eta(\z),\eta(\y)\}$ is small relative to a pre-defined tolerance parameter $\epsilon$. The most basic form of ABC is presented in Algorithm \ref{ABC}. 
\begin{algorithm}
	\caption{ABC Algorithm}\label{ABC}
	\begin{algorithmic}[1]
		\State Simulate ${\theta }^{i}$, $i=1,2,...,N$, from $\pi({\theta }),$
		\State Simulate $\mathbf{z}^{i}=(z_{1}^{i},z_{2}^{i},...,z_{n}^{i})^{\prime }$, $i=1,2,...,N$, from $P_{\theta^i}$;
		\State For each $i=1,...,N$, accept ${\theta }^{i}$ if $d\{\eta(\mathbf{z}^{i}),\eta(\mathbf{y})\}\leq \epsilon$, where $\epsilon$ denotes a user chosen tolerance parameter $\epsilon$. Otherwise, reject ${\theta }^{i}$.
	\end{algorithmic}
\end{algorithm}

ABC thus produces draws of $\mathbf{\theta }$ from an approximation to $\pi(\theta|\y)$ that is no longer conditioned on the full data set $\mathbf{y}$, but on statistics $\mathbf{%
	\eta (y)}$. In what follows, we denote the ABC posterior as $%
\pi_{\epsilon }[{\theta |\eta (\y)}]$, to make the dependence on $\epsilon$ and the statistics $\eta(\y)$ transparent. Throughout, we use the following common representation of the ABC posterior:
\begin{flalign}
\pi_{\epsilon }[{\theta |\eta (\y)}]&=\int_{}\pi_{\epsilon}[\theta,\z|\eta(\y)]\dt\z =\frac{P_{\theta}\left[d\{\eta(\z),\eta(\y)\}\leq \epsilon\right]\pi(\theta)}{\int_\Theta P_{\theta}\left[d\{\eta(\z ),\eta(\y)\}\leq \epsilon\right]\pi(\theta)\dx\theta}\label{ABC_post},
\end{flalign}where
\begin{flalign*}
P_{\theta}\left[d\{\eta(\z),\eta(\y)\}\leq \epsilon\right]:=\int_{}\1\left\{d\{\eta(\z),\eta(\y)\}\leq \epsilon\right\}p_\theta(\z)\dx \z .
\end{flalign*}

In some cases, the output from Algorithm \ref{ABC} can be adjusted to obtain more accurate posterior approximations via a post-processing approach. The most common  post-processing correction is the linear regression adjustment (\citealp{Beaumont2025}); see \citet{blum2018regression} for a review of this common regression post-processing approach. This regression-adjustment version of ABC takes the accepted draws $\{\theta^i\}$ from Algorithm \ref{ABC} and ``adjusts'' them by artificially relating them to $\eta(\mathbf{y})$ through the linear regression model $$\theta^i=\mu+\beta^{\prime} \eta(\mathbf{z}^i)+\nu_i,$$where $v_i$ denotes the model residual. 

Using this regression model, the original $\theta^i$ are adjusted via
$$
\tilde{\theta}^i=\theta^i+\hat{\beta}^{\prime}\{\eta(\mathbf{y})-\eta(\mathbf{z}^{i})\},
$$where $\hat{\beta}$ is most often obtained using weighted least squares. More specifically, for $K_\epsilon(\cdot):=K(\cdot)/\epsilon$, with $K(\cdot)$ some bounded kernel function, and $w_i\propto K_\epsilon(d\{\eta(\z^i),\eta(\y)\})$, $\hat{\beta}$ is obtained by minimizing (in $\mu,\beta$)
$$
\sum_{i=1}^{N}\left(\theta^i-\mu-\beta'\eta(\z^i)\right)^2w_i.
$$
This becomes ordinary least squares when selecting  $K_\epsilon(d\{\eta(\z^i),\eta(\y)\})=\1\left\{d\{\eta(\z^i),\eta(\y)\}\leq \epsilon \right\}$.

\subsection{Model Misspecification in ABC}

ABC implicitly assumes that the model used to generate the simulated summary statistics $\eta(\zb)$ can replicate the behavior of the observed summary statistics $\eta(\y)$. That is, ABC is only required to match those features of the data that are measured by $\eta(\cdot)$. This differs from a standard Bayesian framework based on the likelihood function, where, under general regularity conditions, the posterior ultimately gives higher probability mass to values of $\theta\in\Theta$ that ensure the Kullback-Leibler (KL) divergence
$$\mathcal { D } \left( P_ { 0 } \| P_ { \theta } \right) =\int \log \left\{ \frac { \dx P_ { 0 } ( \mathbf { y } ) } { \dx P_\theta(\y) } \right\} \dx P_ { 0 } ( \mathbf { y } ) $$ is as close to zero as possible. 

In contrast to likelihood-based Bayesian inference, ABC is based on matching simulated and observed summary statistics. Therefore, the meaningful concept of model misspecification in ABC is that the choice of the assumed model, allied with our specific choice of summary statistics, can replicate the observed value of the summary statistic $\eta(\y)$. More formally, let $b_0:=\plim_n \eta(\y)$ and $b(\theta):=\plim_n \eta(\mathbf{z})$, where $\plim_n$ denotes probability limit as $n\rightarrow\infty$.  Using the framework of \cite{marin2014relevant}, and following \cite{frazier2017model}, we formalize this notion of misspecification as follows. 
\begin{definition}\label{def1}
The model $P_\theta\times \Pi$ is ABC misspecified, for the given summary statistic map $\eta(\cdot)$, if $$\inf_{\theta\in\Theta}\|b(\theta)-b_0\|>0.$$
\end{definition}
\noindent Intuitively, ABC misspecification means that, asymptotically, $\eta(\y)$ must be in the range of $\eta(\z)$ for some value of $\theta$, and when $\zb $ is simulated under $P_\theta$. Misspecification as in Definition \ref{def1} has also been referred to as \textit{incompatibility} by \cite{marin2014relevant}. Throughout the remainder we will use the two terms interchangeably. 


As recently discussed by \cite{frazier2017model}, when ABC is based on a model that is misspecified, in the sense of Definition \ref{def1}, the resulting inference can be misleading. More specifically, the ABC posterior has non-standard asymptotic behavior, and regression adjustment ABC can give unreliable results. We briefly illustrate the behavior of ABC under model misspecification using a toy example. 

\begin{example}[Misspecified Normal Model]\label{ex:one}\normalfont
Consider an artificially simple example where the assumed DGP for $z_1,\dots,z_n$ is independent and identically distributed (iid) as $\mathcal{N}(\theta,1)$ but the actual DGP is $y_1,\dots,y_n$ iid from the following mixture of normal random variables $$(2/3)\mathcal{N}(\theta,1)+(1/3)\mathcal{N}(\theta,\sigma^2).  $$ That is, the assumed DGP maintains an incorrect assumption about the class of distributions under analysis. We take as our summary statistics for ABC inference the sample mean and variance, i.e., $\eta(\y)=(\eta_1(\y),\eta_2(\y))'$, where $\eta_1(\y)=\bar{y}$ and $\eta_2(\y)=s^2$. 
 
Consider inference on $\theta$ using two different versions of ABC: the accept/reject approach (hereafter, ABC), where we take $d\{x,y\}=\|x-y\|$ to be the Euclidean norm, {and a local linear regression adjustment approach to ABC (hereafter, ABC-Reg). Following \cite{Beaumont2025}, we take as the kernel function, $K_\epsilon(t)$, the Epanechnikov kernel:
	$K_\epsilon(t)=  c \epsilon ^ { - 1 } \left( 1 - ( t / \epsilon ) ^ { 2 } \right) $, if $t\le\epsilon$, and zero else, where $c$ is a normalizing constant. }

{To demonstrate how these two approaches behave under varying levels of model misspecification}, we fix $\theta=0$ in the true model, and simulate ``observed data'' $\mathbf{y}$ according to different values of $\sigma$. The sample size across the experiments is taken to be $n=100$. We consider a sequence of simulated data
sets for $\y$ such that each corresponds to a different value of
$\sigma$, with $\sigma$ taking values from
${\sigma}^{}=1$ to ${\sigma}^{}=5$ with evenly spaced
increments of $.1$. Across all the data sets we fix the random numbers used to generate
the observed data and only change the value of $\sigma$ to isolate
the impact of model misspecification.

{Our prior beliefs are given by $\theta\sim \mathcal{N}(0,25)$. We implement ABC using $N=1.0\cdot10^6$ simulated pseudo-data sets generated according to $z^{j}_{i}\stackrel{iid}{\sim} \mathcal{N}(\theta^j,1)$.  For both ABC and ABC-Reg, we set $\epsilon$ to be the .05\% quantile of the simulated distances $\|\eta(\mathbf{y})-\eta(\mathbf{z}^{j})\|$. To further isolate the impact of randomness on this procedure, we use the same simulated data across the different observed data sets; i.e., both ABC procedures  use the same simulated data across the different values of $\sigma$. By recycling the same data sets across the experiments, and by controlling the randomness in the observed data, differences in the ABC output across the experiments can be attributed to the changing value of $\sigma$.}

Figure \ref{fig2} plots the ABC and ABC-Reg posteriors across the different values of ${\sigma}^{}$. The results demonstrate that model misspecification {induces dramatic differences between the two ABC approaches}. Indeed, while the posterior mean of ABC remains centred over the pseudo-true value $\theta=0$, the posterior mean of ABC-Reg continually shifts towards smaller values of $\theta$ as the level of misspecification increases. 
\begin{figure}[h!]
	\centering 
	\includegraphics[width=20cm, height=8cm]{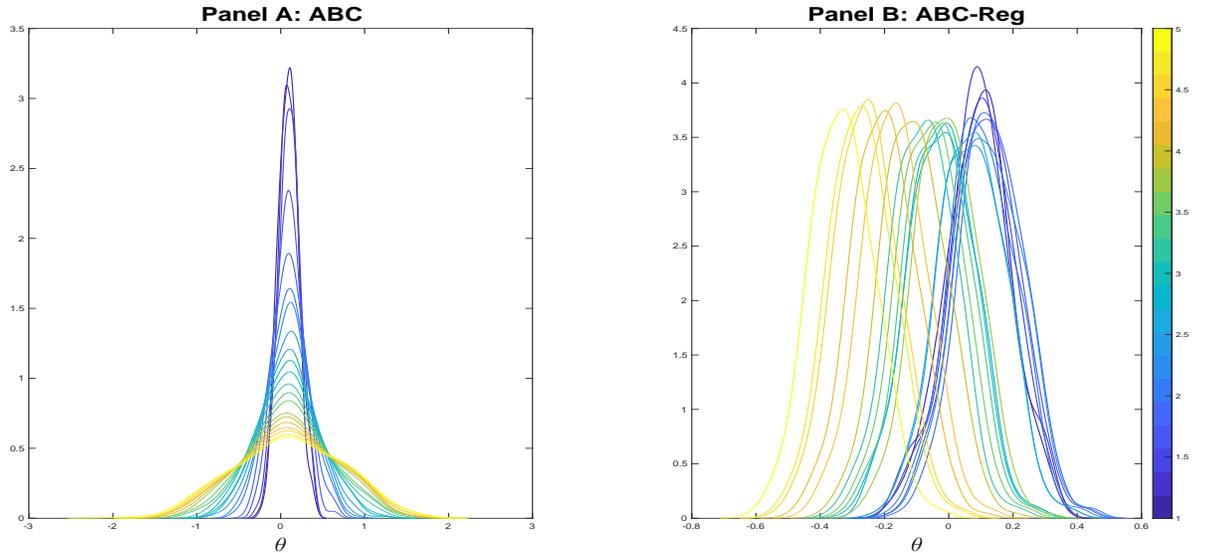} 
	\caption{Comparison of ABC and ABC-Reg posteriors across varying levels of model misspecification. The color of the density represents the level of model misspecification, with $\sigma=1$ encoded as blue, and the color becoming lighter as misspecification increases.} 
	\label{fig2} 
\end{figure}
\end{example}

\section{Robust Approximate Bayesian Computation}
Herein, we propose two possible strategies for conducting robust ABC (R-ABC) inference when the model and summaries may be incompatible (as in Definition \ref{def1}). The first approach augments the simulated summaries with additional free parameters, while the second approach adjusts $\{\eta(\y)-\eta(\z)\}$ directly via the choice of metric used to select draws. Both specifications allow us to conduct reliable inference on the model parameters, and can be used within regression adjustment ABC to produce post-processed ABC output that is well-behaved when the model is misspecified.

\subsection{Summary Adjustment ABC}
ABC misspecification, i.e., model incompatibility, means that, with probability approaching one, $\eta(\y)$ is not in the range of $\eta(\z)$, for any $\theta\in\Theta$. To ensure that $\eta(\y)$ remains within the range under incompatibility, we propose to adjust $\eta(\z)$ by adding a vector of parameters with large prior support. Define this $d_\eta$-dimensional vector of parameters as $\Gamma$, where $\Gamma=(\gamma_1,\dots,\gamma_{d_\eta})^\prime\in\mathcal{G}\subset\mathbb{R}^{d_{\eta}}$, 
 and denote the joint vector of unknown parameters as $\zeta:=(\theta^{\prime},\Gamma^\prime)^\prime\in\Theta\times \mathcal{G} \subset\mathbb{R}^{d_\theta}\times\mathbb{R}^{d_{\eta}}$. We then propose to use as the summary statistics for inference in ABC, the new vector of simulated summaries
\begin{flalign*}
\phi(\z,\Gamma )&=\eta(\z)+\Gamma.
\end{flalign*}
Denoting the prior on $\zeta$ by $\pi(\zeta)$, we define the robust ABC-summary (R-ABC-S) posterior as
\begin{flalign}
\pi_{\epsilon }[{\zeta |\eta (\y)}]&=\int_{\z}\pi_{\epsilon}[\zeta,\z|\eta(\y)]\dx\z =\frac{P_{\zeta}\left[d\{\phi(\z,\Gamma),\eta(\y)\}\leq \epsilon\right]\pi(\zeta)}{\int_\zeta P_{\zeta}\left[d\{\phi(\z,\Gamma),\eta(\y)\}\leq \epsilon\right]\pi(\zeta)\dx\zeta}\label{ABC_post_R},
\end{flalign}where $$P_{\zeta}\left[d\{\phi(\z,\Gamma),\eta(\y)\}\leq \epsilon\right]:=\int_{\z}\1\left[d\{\phi(\z,\Gamma),\eta(\y)\}\leq \epsilon\right]p_\theta(\z)\dx \z.$$ We refer to the posterior in equation \eqref{ABC_post_R} as the R-ABC-S posterior since it employs a version of the summary statistics that are, by construction, always compatible with the observed summaries $\eta(\y)$, and hence robust to model misspecification.

Rejection sampling from the above specification is no more difficult than sampling from the standard ABC target in equation \eqref{ABC_post}: given draws $\zeta^i$ from $\Pi(\zeta)$, the only difference is that we accept the pair $\zeta^i=(\theta^i,\Gamma^i)$ when $d\{\phi(\z,\Gamma),\eta(\y)\}\leq \epsilon$, rather than just $\theta^i$. 

Given an accepted sequence of parameter draws and simulated summary statistics, $\{\zeta^i,\phi(\z^i,\Gamma^i)\}_{i=1}^{M}$, $M\leq N$, a regression adjusted R-ABC-S approach (hereafter, R-ABC-S-Reg) can be implemented using the simulated statistic $\phi(\z^i,\Gamma^i)$. Such an ``adjusted'' version of ABC-Reg produces draws according to  
\begin{flalign*}
\tilde{\theta}^i&=\theta^i+\hat{\beta}^{\prime}\{\eta(\mathbf{y})-\phi(\mathbf{z}^{i},\Gamma^i)\}.
\\\hat{\beta}&= \left[\frac{1}{M_{}}\sum_{i=1}^{M}\left(\phi(\mathbf{z}^{i},\Gamma^i)-\bar{\phi}\right)\left(\phi(\mathbf{z}^{i},\Gamma^i)-\bar{\phi}\right)^{\prime}\right]^{-1}\left[\frac{1}{M_{}}\sum_{i=1}^{N}\left(\phi(\mathbf{z}^{i},\Gamma^i)-\bar{\phi}\right)\left(\theta^i-\bar{\theta}\right)\right].\end{flalign*} where $\bar{\phi}:=\sum_{i=1}^{M}\phi(\mathbf{z}^{i},\Gamma^i)/M$.

Since ABC requires a generative mechanism to sample from the R-ABC-S posterior in equation \eqref{ABC_post_R}, before we complete the discussion on R-ABC-S, we must discuss the choice of priors for the $\Gamma$ components. 

\subsubsection*{Prior Choice: Laplace Priors}
We note that there is no reason to believe \emph{a priori} that $\theta$ and $\Gamma$ are related. Therefore, we take as our overall prior on $\zeta=(\theta',\Gamma')'$ $$\pi(\zeta):=\pi(\theta)\pi(\Gamma).$$In addition, there is also no reason to believe, \emph{a priori}, that the components of $\Gamma$ are dependent, and so we further restrict the prior to have independent elements: $$\pi(\Gamma):=\prod_{j=1}^{d_\eta}\pi(\gamma_j).$$

Given that some components of the original $\eta(\z)$ are likely to be compatible with some components of $\eta(\y)$, we must ensure that the addition of the $\Gamma$ components does not unduly affect the compatible components of $\eta(\z)$. Therefore, we should use a prior that places the vast majority of its mass near the origin.

With these dual requirements in mind,  we propose to follow the Bayesian lasso literature (\citealp{park2008bayesian}) and use independent Laplace (i.e., double-exponential) priors for each component of $\Gamma$, with fixed location $0$ and common scale $\lambda>0$:
\begin{equation}\label{eq:laplace_prior}
\pi(\Gamma):=\prod_{j=1}^{d_\eta}\lambda e^{-\lambda|\gamma_{j}|}=\lambda^{d_\eta}e^{-\lambda\sum_{j=1}^{d_\eta}|\gamma_j|}.
\end{equation} When convenient, we denote this prior by $\text{La}(0,\lambda)$. The Laplace prior for $\Gamma$ guarantees that the majority of the prior mass for $\gamma_j$ is near the origin, but has thick enough tails so that $\phi(\zb,\zeta)$ is compatible with virtually any $\eta(\y)$ that would be used in practice. 

The hyper-parameter $\lambda$ should be chosen in a manner that allows the parameters $\Gamma$ to correct for the existence of  incompatible summaries, when they are in evidence. However, $\lambda$ should also be chosen so that the tails of the summaries are not too thick, i.e., so that the variance of the summaries is not too large. We take as the default choice for the hyper-parameter $\lambda=0.25$. This places most of the prior support between $\pm2$. The default prior La$(0,0.25)$ is used throughout all numerical experiments conducted in the paper.

\begin{example}[Continued: Misspecified Normal Model]\normalfont
We now return to the simple normal motivating example to demonstrate that the R-ABC-S  approach mitigates the issues highlighted in Figure \ref{fig2}. We maintain the same simulation design and use precisely the same simulated data, with the only additional feature required being the generation of $N=1.0\cdot10^6$ simulated random variables $\Gamma\stackrel{iid}{\sim}\text{La}(0,0.25)$.

Figure \ref{fig3} displays the posteriors for R-ABC-S using accept/reject (ABC) and regression adjustment (ABC-Reg). Critically, unlike the results in Figure \ref{fig2}, both approaches are now centered over the pseudo-true value $\theta=0$. Indeed, the posterior ``drift'' that was previously in evidence for ABC-Reg is \textit{no longer} in evidence for R-ABC-S-Reg.   
\begin{figure}[h!]
	\centering 
	\includegraphics[width=18cm, height=8cm]{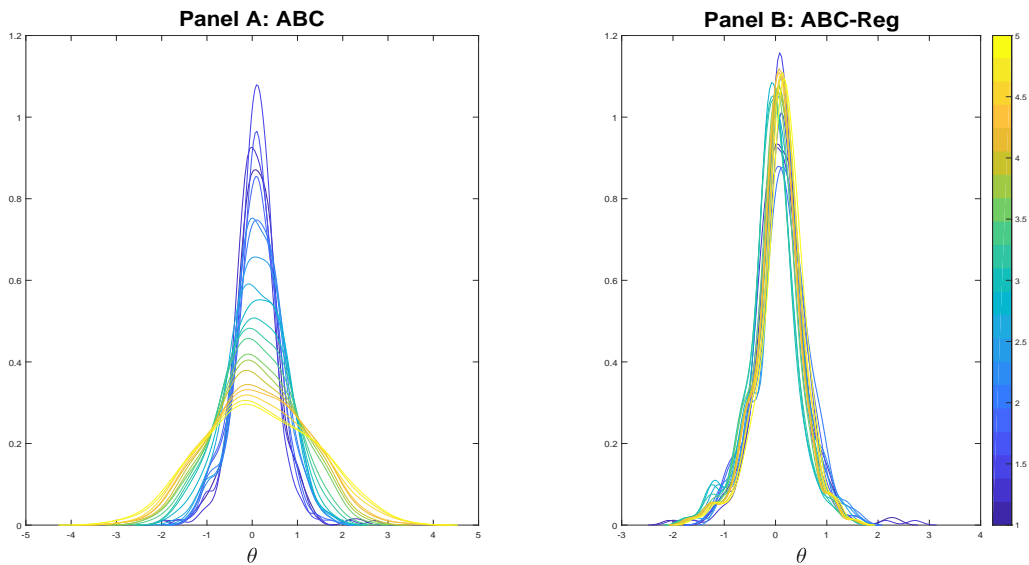} 
	\caption{Comparison of R-ABC-S posteriors for accept/reject ABC (ABC, panel A) and regression adjustment ABC (ABC-Reg, panel B) across varying levels of model misspecification. } 
	\label{fig3} 
\end{figure}

Figure \ref{fig3a} plots the posteriors for the adjustment components obtained from R-ABC-S. The results demonstrate that the $\gamma_1$ adjustment component is indistinguishable from the prior. However, the posterior mean for the $\gamma_2$ component shits further away from the prior mean, of zero, as the level of misspecification increases. This behavior is what allows the R-ABC-S posteriors for $\theta$ to remain centered over the true value $(\theta=0)$ as the level of model misspecification increases. The $\gamma_2$ component directly corresponds to the component of the observed data that we can not match with our assumed model. That is, the behavior of the posterior components for $\gamma$ can be used to detect which of the summary statistics the assumed model can not match.
\begin{figure}[h!]
	\centering 
	\includegraphics[width=18cm, height=8cm]{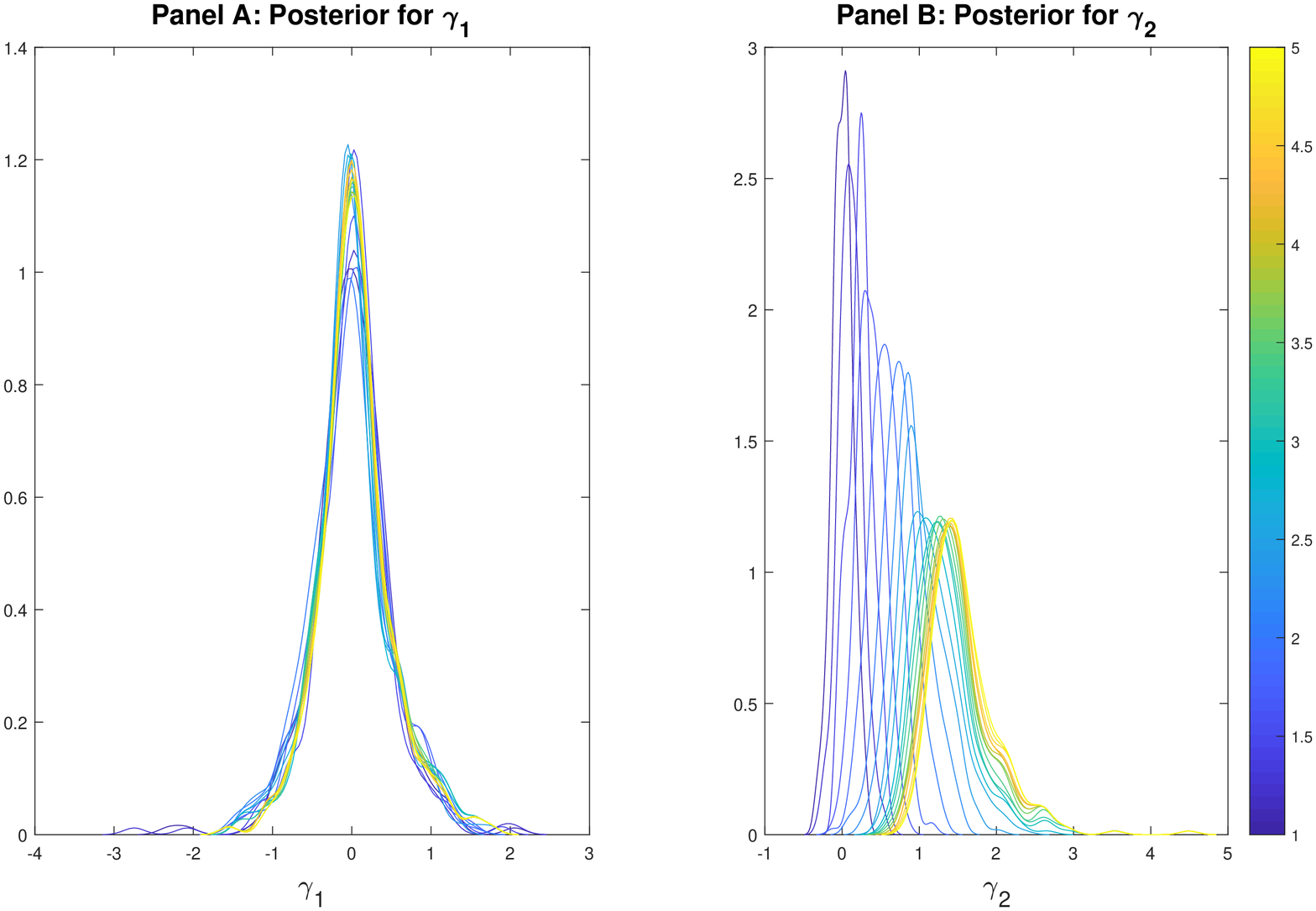} 
	\caption{Posteriors for the adjustment components, $\Gamma=(\gamma_1,\gamma_2)'$, for the R-ABC-S procedure across varying levels of model misspecification. } 
	\label{fig3a} 
\end{figure}

\end{example}

\subsection{Weighted Adjustment ABC}
The second approach we propose is based on scaling the individual summaries so that if $\{\eta(\zb)-\eta(\y)\}$ is large, the scaled version of the summaries can still be made small. Such an approach can be implemented using the vector of summaries
$$
\varphi(\z,\zeta):=\Gamma\odot\{\eta(\y)-\eta(\z)\},
$$where $\odot$ denotes the Hadamard product, and $\Gamma\in\mathcal{G}\subset\mathbb{R}_{+}^{d_\eta}$ is a vector of positive parameters. Given $\varphi(\z,\zeta)$, we then propose to choose values of $\zeta=(\theta^\prime,\Gamma^\prime)^\prime$ in ABC such that $\|\varphi(\z,\zeta)\|\le \epsilon$. Note that this can occur for two reasons: firstly, if $\{\eta(\zb)-\eta(\y)\}$  is small, then $\|\varphi(\z,\zeta)\|$  will remain small even for moderately large $\Gamma$; secondly, if $\{\eta(\zb)-\eta(\y)\}$ is large, the term $\|\varphi(\z,\zeta)\|$ can be made small by taking $\Gamma$ small enough.

For $\|\cdot\|$ denoting the Euclidean norm,  an equivalent interpretation of the above is that we are choosing a weighted norm (with random weights) under which to compare $\{\eta(\y)-\eta(\z)\}$. In particular, for $X\in\mathbb{R}^{d_\eta}$, and $W$ an $(d_\eta\times d_\eta)$-dimensional positive-definite matrix, define the weighted norm $\|X\|_{W}:=\sqrt{X'WX}$. Then, we see that $\|\varphi(\z,\zeta)\|$ is equivalent to 
\begin{flalign}\label{eq:varadj}
\|\varphi(\zb,\zeta)\|^2_{V(\Gamma)}=\{\eta(\y)-\eta(\z)\}'V(\Gamma)\{\eta(\y)-\eta(\z)\},\text{ where } V(\Gamma):=\text{diag}\{
\gamma^2_1,\dots,\gamma^2_{d_\eta}\}.
\end{flalign}

Using the weighted norm $\|\cdot\|_{V(\Gamma)}$ in place of the metric $d\{\cdot,\cdot\}$ in the ABC posterior, and for $\pi(\zeta)$ an appropriate prior on $\zeta$, the robust ABC-Weighted (R-ABC-W) posterior is given as: 
\begin{flalign}
\pi_{\epsilon }[{\zeta |\eta (\y)}]&=\int_{\z}\pi_{\epsilon}[\zeta|\eta(\y)]\dx \z =\frac{P_{\zeta}\left[\|\eta(\y)-\eta(\z)\|_{V(\Gamma)}\leq \epsilon\right]\pi(\zeta)}{\int_\zeta P_{\zeta}\left[\|\eta(\y)-\eta(\z)\|_{V(\Gamma)}\leq \epsilon\right]\pi(\zeta)\dx\zeta}\label{ABC_post_W},
\end{flalign}where
\begin{flalign*}P_{\zeta}\left[\|\eta(\y)-\eta(\z)\|_{V(\Gamma)}\leq \epsilon\right]:=\int_{\z}\1\left\{\|\eta(\y)-\eta(\z)\|_{V(\Gamma)}\leq \epsilon\right\}p_\theta(\z)\dx \z. 
\end{flalign*} 
Similar to R-ABC-S, a regression adjusted version of R-ABC-W (hereafter, R-ABC-W-Reg) can be implemented by replacing the simulated statistic $\eta(\z)$, with the simulated statistic $\varphi(\z,\zeta)$. 

\subsubsection*{Prior Choice: Exponential Prior}
The weighted adjustment (R-ABC-W) requires a prior choice for the $\Gamma$ components. While several prior choices exist for $\Gamma$, following the arguments for the prior choice in R-ABC-S, we need to choose a prior for the components of $\Gamma$ so that there is a moderate amount of prior mass near the origin, and enough mass out in the tails to ensure we can detect incompatible summaries. To this end, we consider independent exponential priors for each component $\gamma_i$, $(i=1,\dots,d_\eta)$, with common rate $\lambda>0$: 
\begin{flalign*}
\pi(\Gamma):=\prod_{i=1}^{d_\eta}\lambda e^{-\lambda \gamma_i}=\lambda^{d_\eta}e^{-\lambda\sum_{i=1}^{d_\eta}\gamma_i}.
\end{flalign*}
The hyper-parameter $\lambda$ should be chosen so as not to over-inflate the variance of the simulated summaries that are compatible, but also to ensure that there is enough mass away from the origin to allow us to meaningfully distinguish between large and small differences between $\eta(\y)$ and $\eta(\zb)$. As a default choice for the prior hyper-parameter, we suggest $\lambda=0.5$. This default choice for the prior is used in all subsequent numerical experiments.\footnote{Several sets of simulation experiments suggest that the results are largely insensitive to the choice of the hyper-parameter $\lambda$.}

This prior is not, strictly speaking, a shrinkage prior, but does yield `shrinkage-like' behavior for summaries that are compatible. That is, for the summaries that are compatible, this additional inflation by $\Gamma$ is unnecessary and we expect that, for appropriate choices of $\lambda$, the addition of this component will not greatly affect the corresponding components in the variance. In contrast, for the summaries that are not compatible, this adjustment term is critical to ensure that the variance of the summaries is large enough to contain the observed summary $\eta(\y)$.

\subsection{R-ABC with a Fixed Weighted Distance}
It has been recognized that, for certain choices of $d\{\cdot,\cdot\}$, if the scales of the summary statistics are different the selection of draws in ABC can be dominated by those summaries with larger scales. To circumvent this issue, a popular choice of distance function in ABC is to use a weighted Euclidean norm with fixed weights: For $
W=\text{diag}\left\{w_1,\dots,w_{d_\eta}\right\},\;w_i\geq0,\text{ for all }i=1,\dots,d_\eta$, ABC can be implemented using the weighted distance 
\begin{flalign}\label{eq:w_norm}
\|\eta(\y)-\eta(\zb)\|_{W}&=\sqrt{\{\eta(\y)-\eta(\zb)\}'W\{\eta(\y)-\eta(\zb)\}}= \sqrt{\sum_{i=1}^{d_\eta}\left(w_i\{\eta(\y)-\eta(\zb)\}\right)^2}.
\end{flalign}

Common choices for the weights $w_i$ include the prior predictive standard deviation, $$w_i:=\text{Var}^{-1/2}_{\pi}\left[\eta_i(\zb)\right].$$ More complicated constructions, where the weights $w_i$ are updated iteratively within either a population Monte Carlo ABC (ABC-PMC) or sequential Monte Carlo ABC (ABC-SMC) approach (we refer to \citealp{prangle2017adapting} for examples of such implementations) are also feasible. Such strategies are particularly useful as they allow the weighting of the summaries to adapt within the ABC procedure so that, wherever we are in the posterior space, the weighted summaries should have similar scale. 

The use of weighted distance functions is also possible within R-ABC. In the case of the R-ABC-S approach, the weights in equation \eqref{eq:w_norm} can simply be replaced with 
$$w_i:=\text{Var}^{-1/2}_{\pi}\left[\phi_i(\zb,\Gamma)\right],\text{ where }\phi_i(\zb,\Gamma)=\eta_i(\zb)+\Gamma_i\text{ for } i=1,\dots,d_\eta.
$$In the case of R-ABC-W, two avenues are available. Firstly, similar to R-ABC-S with a weighted distance function, we could consider as our weights the prior predictive standard deviation of $\Gamma\odot \eta(\zb)$, i.e., 
$$ 
w_i:=\text{Var}^{-1/2}_{\pi}\left[\gamma_i\cdot \eta_i(\zb)\right],\text{ for } i=1,\dots,d_\eta.
$$Secondly, for some diagonal matrix $D^{}=\text{diag}\left\{v_1,\dots,v_{d_\eta}\right\}$, and $v_i\ge0$, for all $i=1,\dots,d_\eta$, we can implement a weighted version of R-ABC-W by choosing the weighted Euclidean norm $\|\eta(\y)-\eta(\zb)\|_{W(\Gamma)}$, where 
$$
W(\Gamma):=D^{1/2}\left[I_{d_\eta}+V(\Gamma)\right]D^{1/2'},\text{ and }V(\Gamma):=\text{diag}\{
\gamma^2_1,\dots,\gamma^2_{d_\eta}\}.
$$This latter choice seems to work particularly well in practice and is the version of R-ABC-W that is used throughout the remainder of our numerical experiments. 
\begin{example}[Continued: Misspecified Normal Model]\normalfont
	We now apply R-ABC-W to the misspecified normal example, where we again use the same data. The R-ABC-W approach requires the generation of $N=1.0\cdot10^6$ simulated random variables for the adjustment components, generated according to the default prior choice. We implement R-ABC using the weighted norm $$W(\Gamma):=\begin{pmatrix}1&0\\0&1\end{pmatrix}+\begin{pmatrix}\gamma_1&0\\0&\gamma_2\end{pmatrix}.$$
	
	Figure \ref{fig4} display the posteriors for the R-ABC-W approach, for both the accept/reject (ABC) and regression adjusted (ABC-Reg) versions. Similar to the case of R-ABC-S, we see that both approaches are centered over the true value $\theta=0$ and R-ABC-W-Reg displays significantly less posterior ``drift'' than ABC-Reg.
	\begin{figure}[h!]
		\centering 
		\includegraphics[width=18cm, height=8cm]{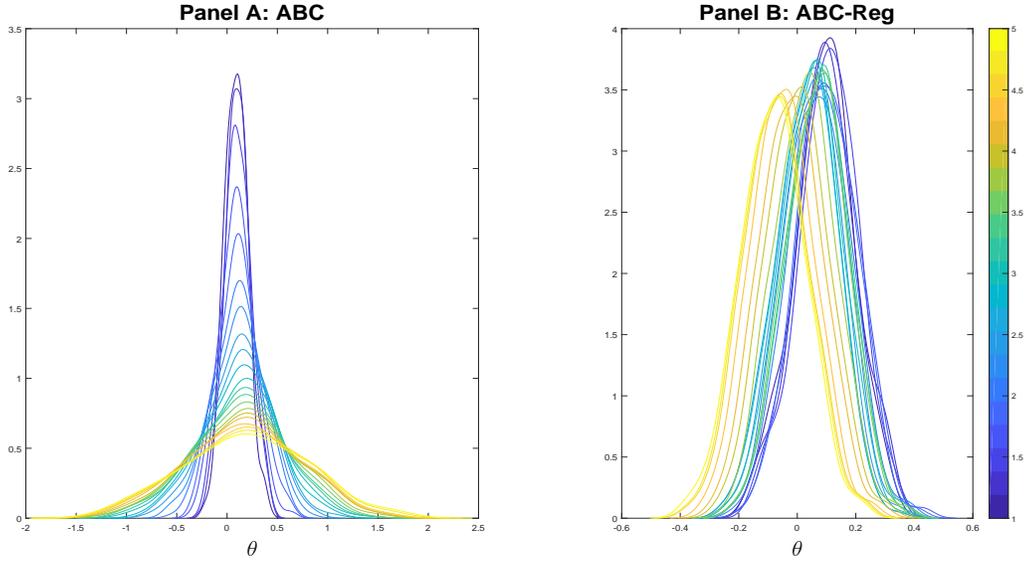} 
		\caption{Comparison of R-ABC-W posteriors for accept/reject ABC (ABC, panel A) and regression adjustment ABC (ABC-Reg, panel B) across varying levels of model misspecification. } 
		\label{fig4} 
	\end{figure}

Figure \ref{fig4a} plots the posteriors for the adjustment components obtained from R-ABC-W. Similar to R-ABC-S, the posteriors for $\gamma_1$ are very similar to the prior. Again, the posterior mean of $\gamma_2$ shifts from the prior mean to accommodate model misspecification.   
\begin{figure}[h!]
	\centering 
		\includegraphics[width=18cm, height=8cm]{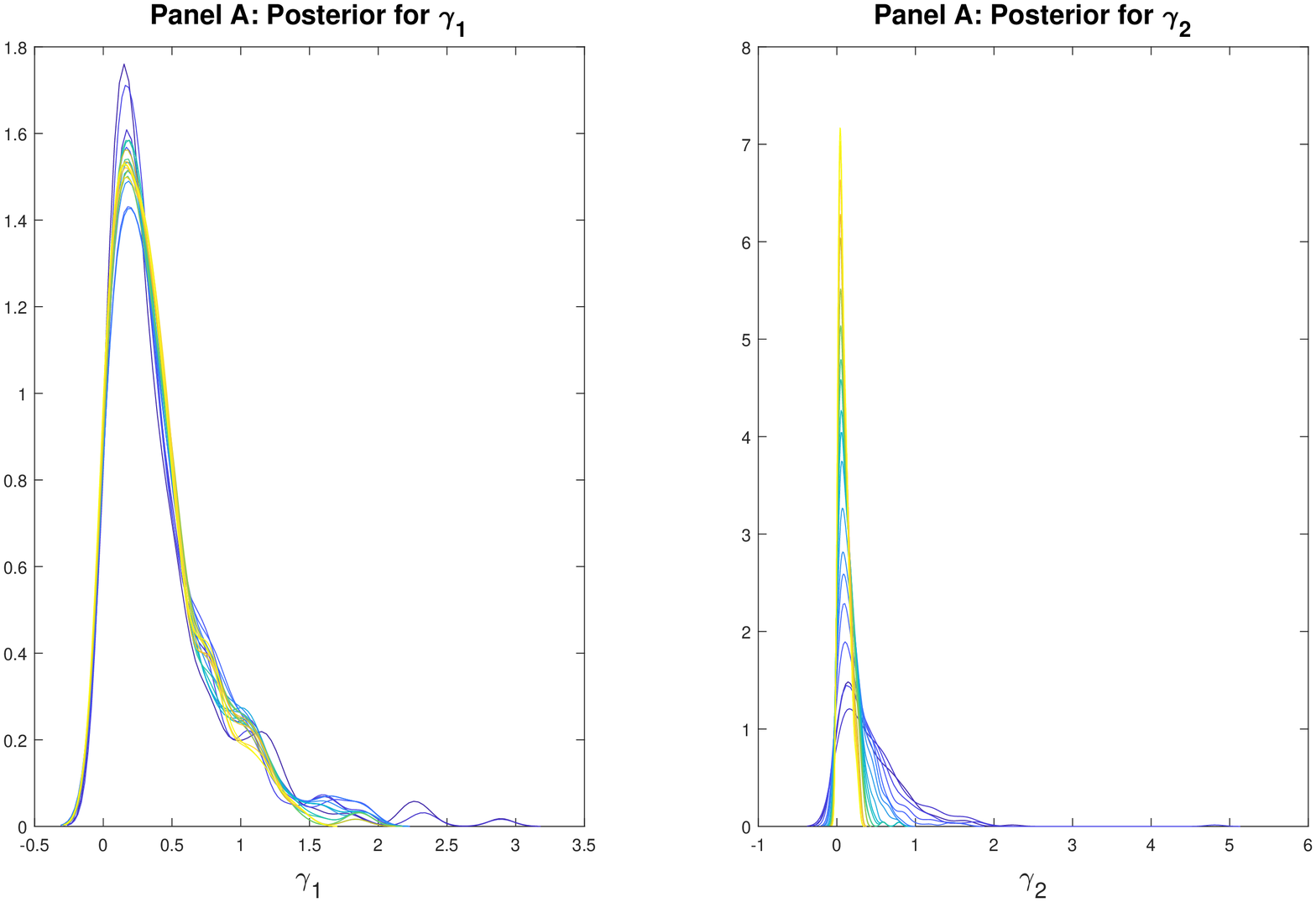} 
	\caption{Posteriors for the adjustment components, $\Gamma=(\gamma_1,\gamma_2)'$, for the R-ABC-W procedure across varying levels of model misspecification. } 
	\label{fig4a} 
\end{figure}

As Figures \ref{fig3} and \ref{fig4} demonstrate, both R-ABC-Reg approaches correct the poor performance of ABC-Reg that is observed at larger levels of model misspecification. However, to further highlight this finding, we graphically compare the posterior means of  vanilla ABC-Reg, R-ABC-S-Reg and R-ABC-W-Reg. In Figure \ref{fig4b}, we plot the posterior means across the different values of $\sigma$. The results emphasize the robustness of R-ABC-Reg relative to vanilla ABC-Reg. 
\begin{figure}[h!]
	\centering 
	\includegraphics[width=18cm, height=9cm]{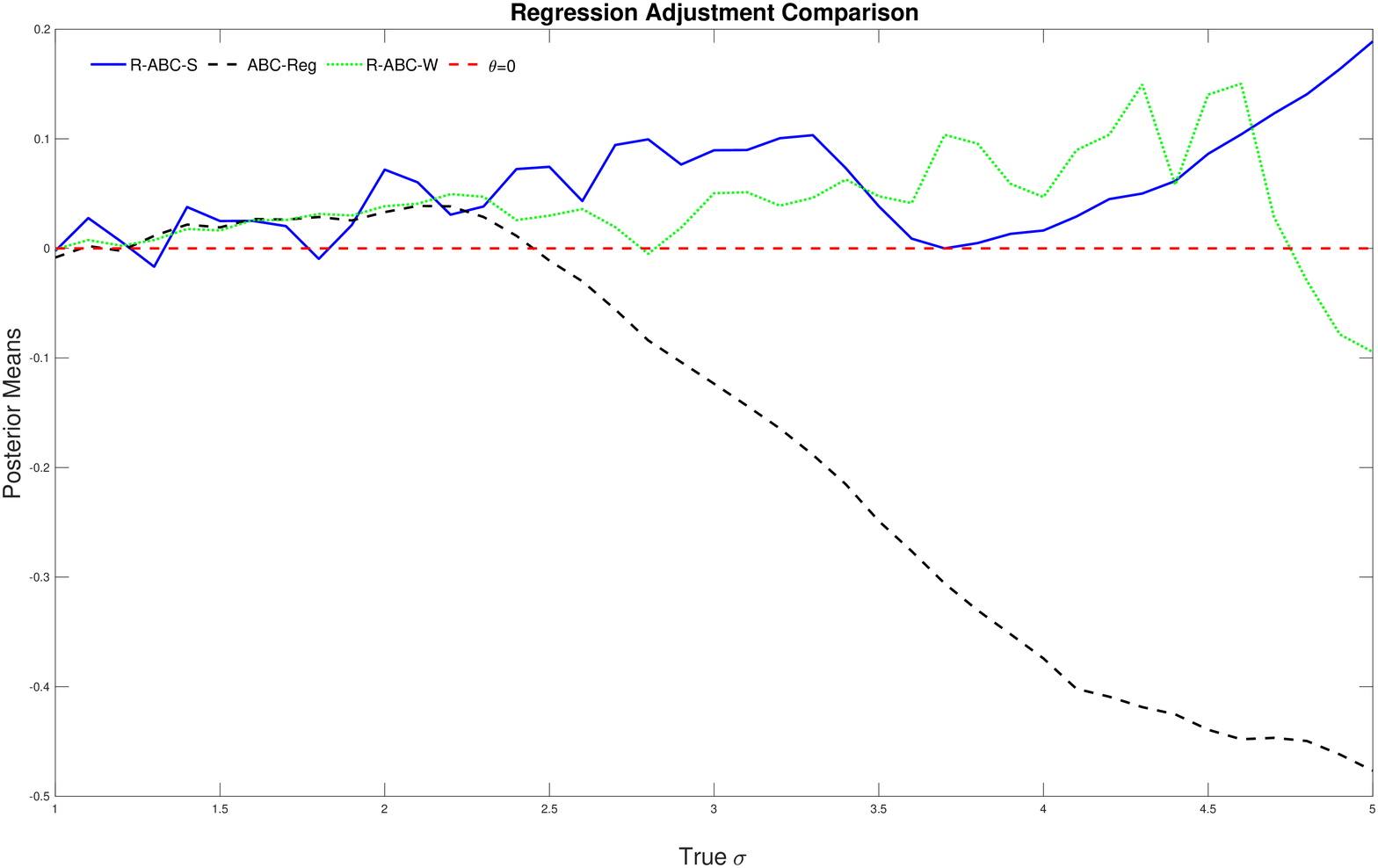} 
	\caption{Posterior means for ABC-Reg, R-ABC-S-Reg and R-ABC-W-Reg, across varying levels of model misspecification. The line $\theta=0$ denotes the true value generating the data.} 
	\label{fig4b} 
\end{figure}

\end{example}

\subsection{Discussion}
The proposed R-ABC approaches rely on two different pathways to deal with model incompatibility. To ensure compatibility can be achieved, R-ABC-S directly adjusts the location of the summaries while R-ABC-W adjusts the scale of the simulated summaries. The difference in these approaches means that  \textit{a priori} there is no reason to prefer one method over the other. Moreover, the simplicity with which both methods can be implemented ensures that there is little cost in applying both procedures in a given application. 

R-ABC is similar to the approach proposed in \cite{FD2019} to deal with model misspecification when inference is conducted using Bayesian synthetic likelihood (\citealp{wood2010}, \citealp{price2018bayesian}). While similar in nature, the motivation behind the two approaches is quite distinct. In particular, \cite{FD2019} were motivated by finding that in misspecified models the BSL posterior can display poor behavior, such as bi-modality and a lack of posterior concentration, and can be difficult to sample using Markov chain Monte Carlo.\footnote{The sampling issues can be traced back to the fact that the synthetic likelihood often displays large variance if the model is misspecified, which ultimately results in low acceptance rates.} In comparison, and as demonstrated theoretically in \cite{frazier2017model},  the accept/reject ABC posterior still displays posterior concentration even under extreme levels of model misspecification, albeit with unreliable uncertainty quantification, whereas the ABC-Reg posterior can display poor behavior. Given this, the benefits of R-ABC are most significant when R-ABC is applied in conjunction with regression adjustment ABC. From this viewpoint, we argue that although the methodology used in this paper and \cite{FD2019} are similar, the reasoning underlying application of R-ABC is entirely distinct from that encountered in the BSL literature. Further, BSL requires that, pointwise in $\theta$, the model summary statistic distribution is regular enough so that a Gaussian assumption is reasonable.   In applications where this assumption is violated, R-ABC may be the preferred approach to robust likelihood-free Bayesian inference.

In the following sequence of examples, we further demonstrate that using R-ABC in conjunction with regression adjustment, can lead to good performance in misspecified models.

\section{Examples}\label{sec:examples}
In this section,  we give further evidence of the significantly improved inferences that can be achieved with R-ABC when the model is misspecified. We also demonstrate that the $\Gamma$ components of the R-ABC approaches can be used to reliably detect which of the summaries may be incompatible with the assumed model.

\subsection{An Additional Normal Example}\label{sec:examplesN}

In this section, we compare the performance of ABC with and without the regression correction, and R-ABC under model misspecification in a toy normal example. Our goal is inference on $\theta$ in the model $$y_i = \theta+v_i,\;v_i  \stackrel{iid}{\sim} \mathcal{N}(0,\sigma^2 ),$$ where we explicitly assume that $\sigma=1$, and generate data in ABC according to
$$z_i = \theta+v_i,\;{v}_i\stackrel{iid}{\sim}\mathcal{N}(0,1).$$ We consider as summary statistics the sample mean $\eta_{1}(\mathbf{y})=\frac{1}{n}\sum_{i=1}^{n}{y}_{i}$ and the sample variance $\eta_{2}(\mathbf{y})=\frac{1}{n-1}\sum_{i=1}^{n}({y}_{i}-\eta_{1}(\mathbf{y}))^{2}$. The sample size across all experiments is taken to be $n=100$, and the prior is $\theta\sim \mathcal{N}(0,5^2)$. 

For this experiment, we generate simulated data
sets for $\y$ under the true model with $\theta=1$, and where each data set corresponds to a different value of
${\sigma}^{}$, with ${\sigma}^{}$ taking values from
${\sigma}^{}=0.5$ to ${\sigma}^{}=5$ with evenly spaced
increments of $0.1$. Across all the data sets we fix the random numbers used to generate
the ``observed'' data and only change the value of ${\sigma}^{}$ to isolate
the impact of model misspecification.
Each ABC approach is based on $N=1.0\cdot10^6$ simulated data sets from the prior predictive distribution, and we retain the draws that lead to the smallest 0.05\% of the simulated distances. 

\begin{sidewaysfigure}[]	
	\centering
	\includegraphics[width=25cm, height=15cm]{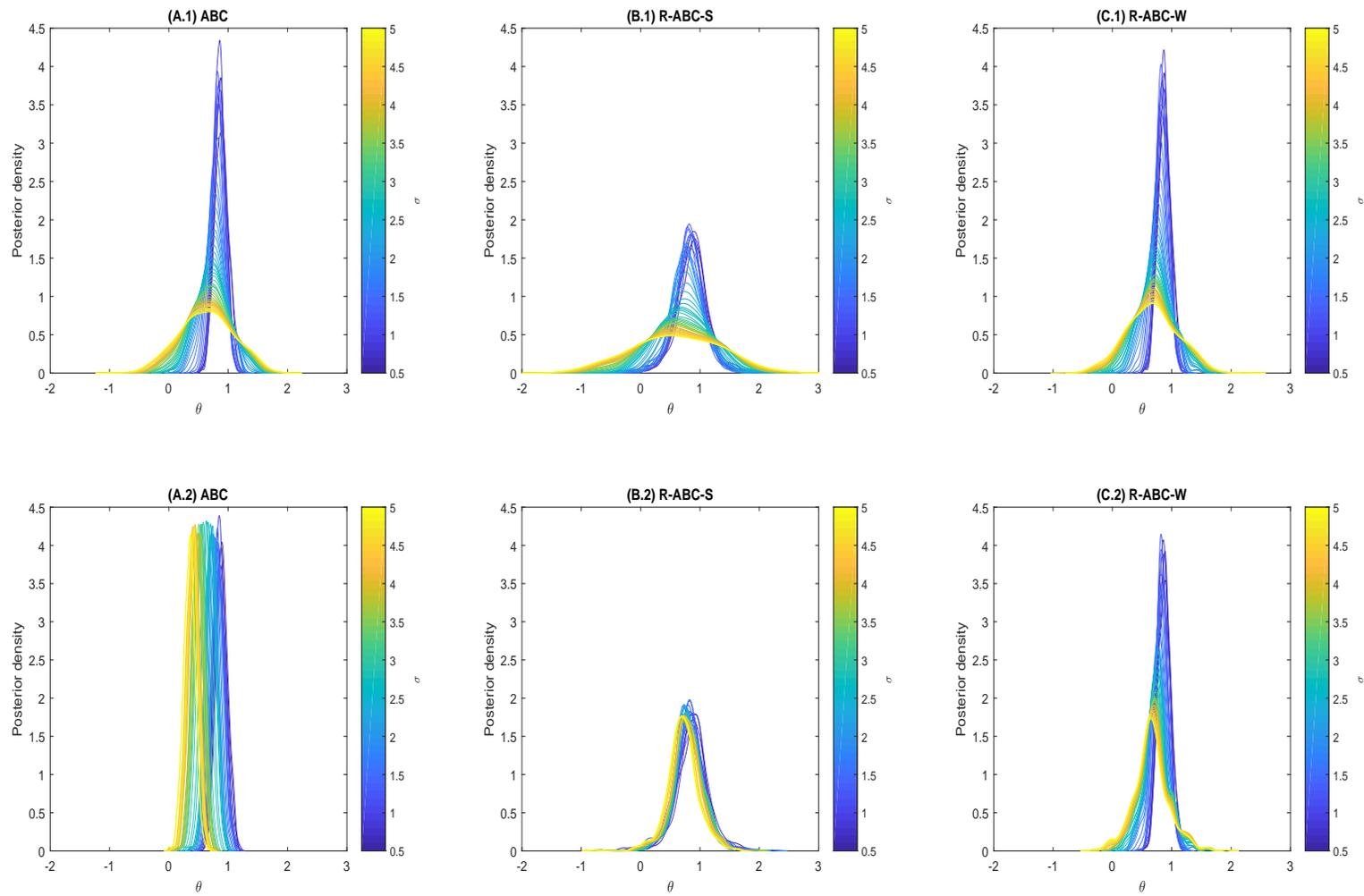}  
	\caption{The top panels of the figure correspond to standard accept/reject ABC posteriors, while the bottom panels are those based on the regression corrected ABC.}%
	\label{fig:normal_rbsl_t}
\end{sidewaysfigure}
\begin{figure}[H]	
	\centering
	\includegraphics[width=17cm, height=13cm]{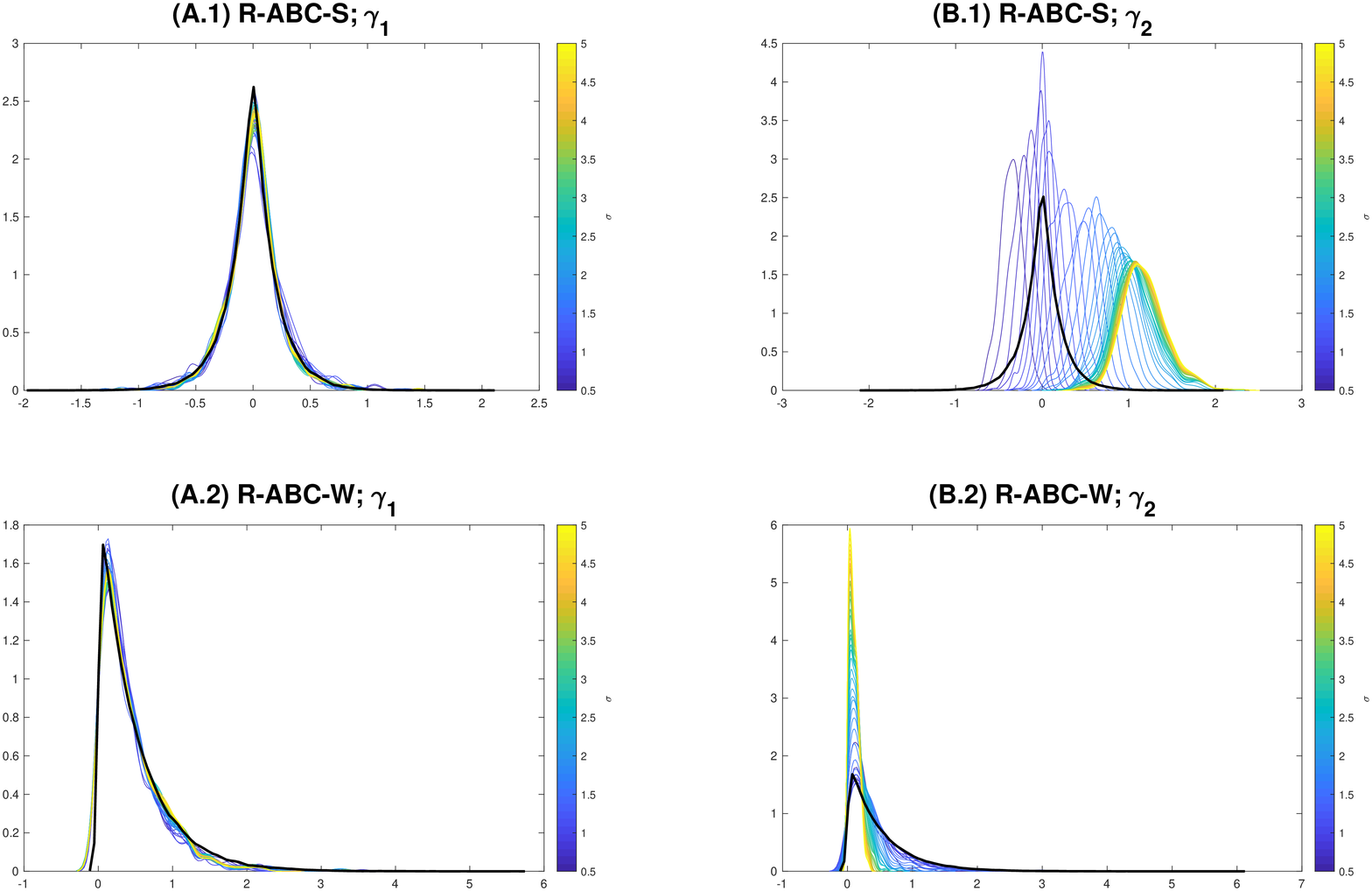}  
	\caption{Marginal posteriors for the adjustment components. The top panels correspond to the posterior for the mean adjustment, while the bottom panels correspond to the adjustment components for the weighted ABC approach. The first panel on the top and the bottom correspond to the adjustment for the first summary and the second panel for the second summary.}%
	\label{fig:normal_rbsl_g}
\end{figure}

We present the results of the experiment graphically in Figures \ref{fig:normal_rbsl_t} and \ref{fig:normal_rbsl_g}. Figure \ref{fig:normal_rbsl_t} presents the posteriors for $\theta$ over the different data sets for each of the six different inference approaches: standard accept/reject ABC, ABC with regression adjustment (ABC-Reg), both methods where we use the summary statistic adjustment (R-ABC-S), as well as both ABC methods where we use the weighted adjustment (R-ABC-W). Analyzing the ABC-Reg posterior in panel (A.2), we see that the ABC-Reg posterior shifts away from the actual mean of the observed data $(\theta=1)$ at large levels of model misspecification. However, the results in panels (B.2) and (C.3) demonstrate that both R-ABC-Reg posteriors do not display this behavior and are (mostly) centered over the true value $\theta=1$. 

Figure \ref{fig:normal_rbsl_g} displays the resulting posterior densities for the $\Gamma$ components across the two adjustment procedures, and across all levels of misspecification. The top panels consider the adjustment components for the mean approach and the bottom panels give the posteriors for the weighted approach. The top and bottom panels are further broken down according to the summary statistics, with the first panels corresponding for the mean summary and the second the variance summary. For comparison purposes, the black figure in each panel represents the prior densities. 

Focusing on the top panel first, we see that the posterior densities for the $\gamma_1$ component, which captures our ability to match the first observed statistic (the mean), are indistinguishable from the prior across all the observed data sets, which implies that we can match the mean of this model regardless of model misspecification. In contrast, in Panel B we see that the second component, which captures our ability to capture the second observed statistics (the variance), looks nothing like the prior, except perhaps at low levels of misspecification.

The above results are not a function of any given sample but are persistent across different samples. To demonstrate this feature we consider a repeated sampling version of the above Monte Carlo experiment, but only for three separate values of $\sigma^2$; namely,  $\sigma^2=1,2,3$. For each of the three values, we simulate 500 replications of the observed data and run the six different ABC procedures across each of the different data sets. For each replication, we record the bias of the posterior mean, the posterior standard deviation, and across the replications we calculate the Monte Carlo coverage of each procedure. The results across the different procedures are tabulated in Table \ref{tab:newTable}. 

The results in Table \ref{tab:newTable} demonstrate that standard regression adjustment ABC (ABC-Reg) has very poor coverage at higher levels of model misspecification. However, we see that the R-ABC-Reg posteriors display much more reasonable coverage. The bias of the posterior mean is roughly similar across ABC and the accept/reject R-ABC approaches, while ABC-Reg and R-ABC-S-Reg have biases that are larger than the other approaches. In comparison, R-ABC-W-Reg has the smallest bias across the methods and across all levels of model misspecification.

Comparing posterior variability across the different methods, we see that the rejection-based R-ABC-S approach displays the largest posterior uncertainty, while ABC-Reg displays the smallest posterior uncertainty. The rejection-based R-ABC-W approach displays results that are similar to ABC, while R-ABC-W-Reg yields posterior uncertainties that are smaller than ABC and R-ABC-S, but which are larger than ABC-Reg. 

Overall, given the relatively small posterior uncertainty, and the well-centred nature of the posterior, we argue that the R-ABC-W-Reg approach performs best across the different levels of model misspecification. 

\begin{table}[h]
\caption{{Monte Carlo coverage (Cov), credible set length (Length), and posterior standard deviation (Std) for the simple normal example under various levels of model misspecification. Cov is the percentage of times that the 95\% credible set contained $\theta^*=1$. Length is the average length of the credible set across the Monte Carlo trials. Std is the average posterior standard deviation across the Monte Carlo trials.\newline }}
\centering%
	\begin{tabular}{lrrrrrrrrrr}
	&   \multicolumn{1}{c}{ABC}     & &       &       &   \multicolumn{1}{c}{R-ABC-S}    &  &  & \multicolumn{1}{c}{R-ABC-W} &  &\\
	& \multicolumn{1}{c}{Cov} & \multicolumn{1}{c}{Bias} & \multicolumn{1}{c}{Std} &       & \multicolumn{1}{c}{Cov} & \multicolumn{1}{c}{Bias} & \multicolumn{1}{c}{Std} &\multicolumn{1}{c}{Cov} & \multicolumn{1}{c}{Bias} & \multicolumn{1}{c}{Std}\\
	$\sigma^2=1$    & 95\% & -.010 & 0.102 &       & 100\%     & -.023  &  0.265 &        94\%&   -.009 &   0.102	\\
	$\sigma^2=2$    & 98\% & -.017 & 0.199 &       & 100\%     &-.023 &   0.292 &        98\%&    -.019 &   0.199	\\
	$\sigma^2=3$   & 	100\%     & -.028 & 0.310 &       & 100\%     & -.042 &   0.446 &        100\%     & -.032 &   0.293	\\
	&       &       &       &       &       &       &  & & &\\
	&   \multicolumn{1}{c}{ABC-Reg}    &  &       &       &  \multicolumn{1}{c}{R-ABC-S-Reg}     &  &  &  \multicolumn{1}{c}{R-ABC-W-Reg}&&\\
	& \multicolumn{1}{c}{Cov} & \multicolumn{1}{c}{Bias} & \multicolumn{1}{c}{Std} &       & \multicolumn{1}{c}{Cov} & \multicolumn{1}{c}{Bias} & \multicolumn{1}{c}{Std} &\multicolumn{1}{c}{Cov} & \multicolumn{1}{c}{Bias} & \multicolumn{1}{c}{Std}\\
	$\sigma^2=1$    & 95\% & -.010 & 0.101 &       & 100\%     & -.023  &  0.264 &        95\%  &  -.009   & 0.101\\
	$\sigma^2=2$   &72\%  & -.075 & 0.098 &       & 100\%     &-.024 &   0.264 &        92\%  &  -.023  &   0.132\\
	$\sigma^2=3$    & 61\%  & -.143 &   0.099 &       & 99\%     & -.088  &  0.266 &       95\%  &  -.011 &   0.167\\
\end{tabular}%
	\label{tab:newTable}%
\end{table}

\subsection{Moving Average Model}
 
To further illustrate R-ABC, we turn to a common toy examples encountered in the approximate inference literature, the moving average model of order two MA(2). Assume  the researcher believes $\mathbf{y}$ is generated according to an MA($q$) model:
\begin{equation*}
z_{t}=e_{t}+\sum_{i=1}^{q}\theta _{i}e_{t-i}, 
\end{equation*}%
where, say, $e_{t}\stackrel{iid}{\sim} \mathcal{N}(0,1)$ and $\theta _{1},...,\theta _{q}$ are such that the roots of the polynomial
\begin{flalign*}
p(x)=1-\sum_{i=1}^{q}\theta _{i}x^{i}
\end{flalign*}all lie outside the unit circle. Specializing this model to the case where $q=2$, we have that 
\begin{equation}
z_{t}=e_{t}+\theta _{1}e_{t-1}+\theta _{2}e_{t-2},  \label{MA2}
\end{equation}and the unknown parameters $\theta=(\theta_{1},\theta_{2})^\prime$ are assumed to obey
\begin{equation}
-2<\theta _{1}<2,\;\theta _{1}+\theta _{2}>-1,\theta _{1}-\theta _{2}<1.
\label{const1}
\end{equation}
Our prior information on $\theta=(\theta_{1},\theta_{2})^\prime$ is uniform over the invertibility region in \eqref{const1}. A useful choice of summary statistics for the MA(2) model are the
sample autocovariances $\eta _{j}(\mathbf{z})=\frac{1}{T}%
\sum_{t=1+j}^{T}z_{t}z_{t-j}$, for $j=0,1,2$. Letting $\eta(\mathbf{z})$ denote the summaries $\mathbf{\eta }\left( \mathbf{z}\right) =(\eta _{0}\mathbf{(z)}, \eta _{1}\mathbf{(z)}, \eta _{2}\mathbf{(z)})^\prime$ and define their probability limit as $b(\theta):=\plim \eta(\zb)$. Then, under the DGP in equations \eqref{MA2}-\eqref{const1}, the limit map $\theta\mapsto b(\theta)$ is given by $$b(\theta)=\begin{pmatrix}
1+\theta^2_{1}+\theta^2_{2},&\theta_{1}(1+\theta_{2}),&\theta_{2}
\end{pmatrix}^{\prime}.$$

Since we are interested in examining the ability of ABC and R-ABC to deal with model incompatibility, we consider that, while the researcher believes the data is generated according to an MA(2) model, equation \eqref{MA2}, the actual DGP for $\mathbf{y}$ evolves according to the stochastic volatility (SV) model  
\begin{flalign}
y_{t}&=\exp(h_{t}/2)u_{t}\nonumber\\h_{t}&=\omega+\rho h_{t-1}+v_{t}\sigma_{v}\label{trueDGP}
\end{flalign} where $0<\rho<1$, $0<\sigma_{v}<1$, $u_{t}$ and  $v_{t}$ and both iid standard Gaussian. In this case, if one takes $\mathbf{\eta }\left( \mathbf{y}\right) :=(\eta _{0}\mathbf{(y)}, \eta_{1}\mathbf{(y)}, \eta _{2}\mathbf{(y)})^\prime$ it follows that, under the DGP in \eqref{trueDGP},
\begin{equation}\label{eq:ma_incomp}
\eta(\mathbf{y})\stackrel{P}{\rightarrow} b_{0}:=\begin{pmatrix}\exp\left( \frac{\omega}{1-\rho}+\frac{1}{2}\frac{\sigma_v^2}{1-\rho^2}\right)
,&0,&0
\end{pmatrix}^{\prime}.\end{equation}
For any value of $\omega,\sigma_v$ and $\rho$ such that $$\exp\left( \frac{\omega}{1-\rho}+\frac{1}{2}\frac{\sigma_v^2}{1-\rho^2}\right)\neq1,$$ it follows that $(P^n_\theta\times\Pi,\eta)$ is not compatible. From the definition of $b(\theta)$ and $b_0$, it also follows that the value that minimizes $\|b(\theta)-b_0\|$ is $\theta^*=(0,0)^\prime$, and is the value the ABC posterior will concentrate onto in the limit.

To understand how ABC and R-ABC perform in this
misspecified model, we consider the following Monte Carlo analysis: we generate $n$ = 1000 observations
from the SV model in (9) and use ABC, ABC-Reg, and R-ABC-S(W) to conduct inference on the unknown parameters in the misspecified
MA(2) model. Each ABC approach uses $N$ = $1.0\cdot10^6$ simulated data sets from the prior predictive
distribution, and accepts values of the parameters that lead to simulated distances in the
smallest .05\% of the overall simulated distances. Both R-ABC-S and R-ABC-W use the default priors, with R-ABC-W again using an identify weighting matrix.

\begin{figure}[h]
	\centering
	\includegraphics[width=16cm, height=8cm]{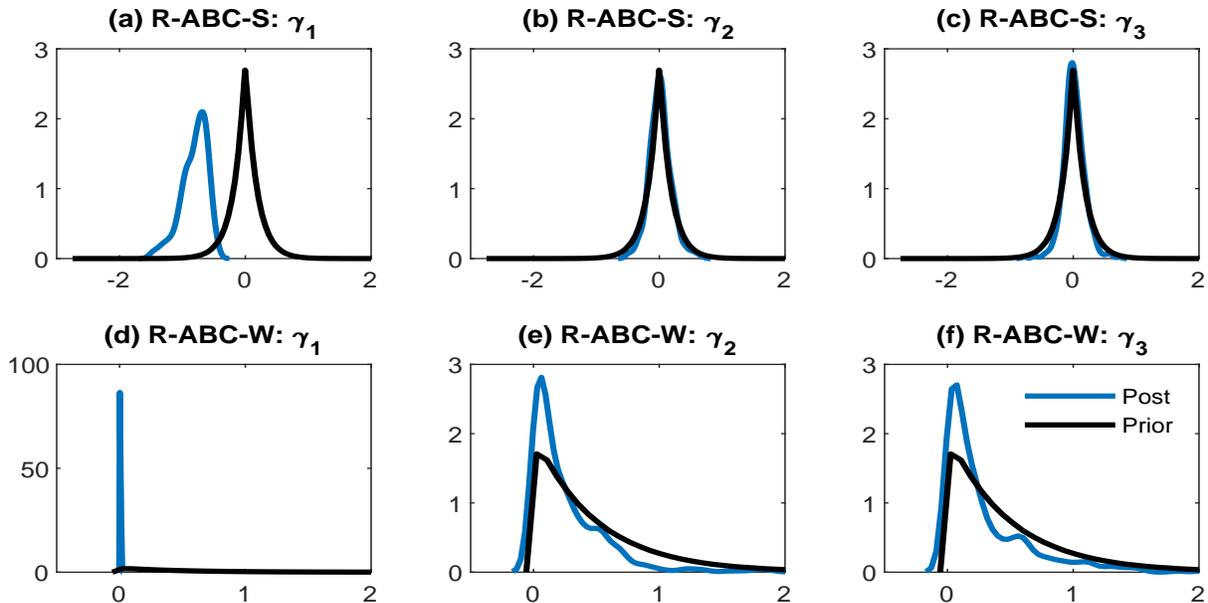} 
	 \caption{Panels (a), (b) and (c) give the R-ABC-S posteriors for the $\gamma_{1},\gamma_2,\gamma_3$, while panels (d), (e) and (f) give the R-ABC-W posteriors for the same components. The priors for the top three components are all $\text{La}(0,\lambda=0.25)$, while the priors for the bottom three panels are $\text{Exp}(0.5)$. }%
	\label{fig:ma2_adjust}
\end{figure}

Given the form of incompatibility in \eqref{eq:ma_incomp}, we would expect that the R-ABC procedures to detect incompatibility in the first summary statistic, the sample variance. Figure \ref{fig:ma2_adjust} displays the results of the adjustment components for the R-ABC-S and R-ABC-W procedures. The results are largely as expected, with the R-ABC procedures clearly detecting that we cannot match the first summary statistic.

\begin{figure}[h]
	\centering
	\includegraphics[width=17cm, height=9cm]{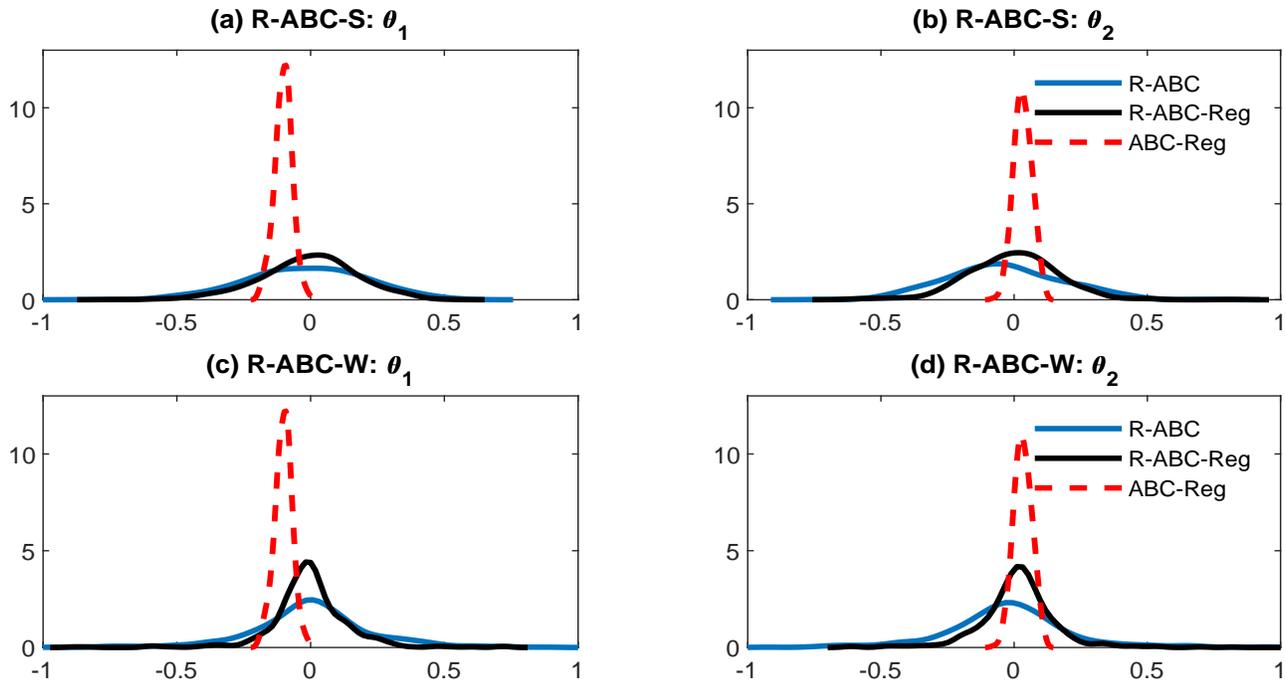} 
	\caption{Panels (a) and (b) provide the R-ABC-S posterior densities for the $\theta_{1}$ and $\theta_2$, while panels (c) and (d) present the R-ABC-W posterior densities for the same components. The blue and black lines indicate the robust ABC posteriors with and without regression adjustment, respectively. The red dashed line displays the non-robust ABC posterior with regression adjustment}%
	\label{fig:ma2_adjust_theta}
\end{figure}

In Figure \ref{fig:ma2_adjust_theta}, we plot the resulting posteriors for the $\theta$ components from the MA(2) model across the different ABC procedures. Similar to the simple normal example, we see that there is substantial differences between the adjusted and unadjusted ABC-Reg posteriors.\footnote{As in the simple normal example, the accept/reject R-ABC and ABC posteriors are similar, and so we do not plot the ABC posteriors in Figure \ref{fig:ma2_adjust_theta} to enhance the readability of the figure.} Most notably, the R-ABC-Reg approaches display larger posterior variability than ABC-Reg and are more closely centered over the pseudo-true value $\theta^*=(0,0)'$. 

Using the same Monte Carlo specification as above, we conduct a repeated sampling exercise that compares the six different ABC procedures. We consider 100 replications and for each replication we record the bias of the posterior mean, the posterior standard deviation, and across the replications we calculate the Monte Carlo coverage of each procedure. We report the results for $\theta_1$ and $\theta_2$ in Table \ref{tab:newTable1}.

The results in Table \ref{tab:newTable1} demonstrates that ABC-Reg has very poor coverage at higher levels of model misspecification.  Indeed, for $\theta_2$, credible sets obtained by regression adjustment do not contain the pseudo-true value in any of the Monte Carlo replications. However, we see that both R-ABC-Reg procedures correct this behavior and deliver posteriors that have more reasonable coverage.  

The results for R-ABC-W-Reg echo those obtained in Section \ref{sec:examplesN}, with R-ABC-W-Reg displaying posterior means with small bias, relatively small standard deviations and reasonable coverage rates. Indeed, R-ABC-W-Reg arguably displays the best behavior across the different methods.

\begin{table}[h]
	\caption{{Bias of the posterior mean (Bias), Monte Carlo coverage (Cov), and posterior standard deviation (Std) for $\theta=(\theta_1,\theta_2)'$ in the MA(2) example. Cov is the percentage of times that the marginal 95\% credible set contained $\theta^*_j=0$, for $j=1,2$. Std is the average posterior standard deviation across the Monte Carlo trials.\newline }}
	\centering%
	\begin{tabular}{lrrrrrrrrrr}
	&   \multicolumn{1}{c}{ABC}     & &       &       &   \multicolumn{1}{c}{R-ABC-S}    &  &  & \multicolumn{1}{c}{R-ABC-W} &  &\\
	& \multicolumn{1}{c}{Cov} & \multicolumn{1}{c}{Bias} & \multicolumn{1}{c}{Std} &       & \multicolumn{1}{c}{Cov} & \multicolumn{1}{c}{Bias} & \multicolumn{1}{c}{Std} &\multicolumn{1}{c}{Cov} & \multicolumn{1}{c}{Bias} & \multicolumn{1}{c}{Std}\\
	$\theta_1$    & 100\%     & -0.0007 & 0.0888 &       & 100\%   & 0.0100 & 0.2249 &        100\%      &  -0.0005 &   0.0979	\\
	$\theta_2$    & 100\%     & -0.0014 & 0.0898 &       & 100\%      & -0.0077 & 0.2034 &        100\%      & 0.0001 &   0.0952 	\\
	&       &       &       &       &       &       &  & & &\\
	&   \multicolumn{1}{c}{ABC-Reg}    &  &       &       &  \multicolumn{1}{c}{R-ABC-S-Reg}     &  &  &  \multicolumn{1}{c}{R-ABC-W-Reg}&&\\
	& \multicolumn{1}{c}{Cov} & \multicolumn{1}{c}{Bias} & \multicolumn{1}{c}{Std} &       & \multicolumn{1}{c}{Cov} & \multicolumn{1}{c}{Bias} & \multicolumn{1}{c}{Std} &\multicolumn{1}{c}{Cov} & \multicolumn{1}{c}{Bias} & \multicolumn{1}{c}{Std}\\
	$\theta_1$    & 0\%     & -0.0028 & 0.0340 &       & 100\%     & 0.0297 & 0.193 &        100\%      & 0.0037 &   0.0537\\
	$\theta_2$   &100\%     & -0.0040 & 0.0339 &       & 100\%      & 0.0037 & 0.1691 &        100\%     & 0.0001 & 0.0512\\
\end{tabular}%
\label{tab:newTable1}%
\end{table}

\subsection{$\alpha$-Stable Stochastic Volatility Model}
We now apply the R-ABC-S approach to conduct inference on the behavior of daily log-returns on the S\&P500 index using data from 2 January 2013 until 7 February 2017, which yields 1033 daily observations. Returns are calculated using open-to-close daily prices. The return series is standardized by dividing each observation by the standard deviation calculated over the length of the series and then subtracting the overall mean. 

Following several authors, including \cite{carr2003finite}, \cite{LC2009} and \cite{martin2019auxiliary}, we consider that $r_t$ is generated according to the following stochastic volatility model:
\begin{flalign*}
	r_{t}&=\sigma_{t} w_{t}, \\ \ln \sigma^2_{t}&=\theta_{1}+\theta_{2} \ln \sigma^2_{t-1}+\theta_{3} v_{t}.
\end{flalign*}The error term $w_t$ is assumed to be an iid increment from an $\alpha$-stable Levy process, $w_t\stackrel{iid}{\sim} \mathcal{S}(\theta_4,\theta_5,0,1)$, with tail index $\theta_4\in(1,2]$, skewness parameter $\theta_5\in[-1,1]$, location zero and unit scale. The error term $v_t$ is assumed to be iid Gaussian. Due to the heavy tailed nature of the error terms, the $\alpha$-stable volatility process can capture the high levels of return volatility that can exist in many series. As such, this model is particularly well-suited for analyzing returns on volatile stocks and stock indices.

To simplify the analysis, in this example we fix the skewness parameter, $\theta_5$, and the volatility location parameter, $\theta_1$, to both be zero. We consider the following priors over the remaining parameters:
$$
\theta_2\sim\mathcal{U}(.7,1),\;\;\theta_3\sim\mathcal{U}(0.001,0.50),\;\;\theta_4\sim\mathcal{U}(1.2,2).
$$ These specific priors were also used in \cite{martin2019auxiliary} and reflect existing empirical evidence on the support of these parameters based on previous studies of daily returns on the S\&P500 index.

We generate summary statistics for ABC inference using an auxiliary model that can cater for the heavy tailed nature of the observed data. In particular, we consider an auxiliary model based on a first-order generalized autoregressive conditional heteroscedastic (GARCH(1,1)) model
\begin{flalign*}
r_{t}&=x_{t} \epsilon_{t} \\ x_{t}&=\beta_{1}+\beta_{2} x_{t-1}\left|\epsilon_{t-1}\right|+\beta_{3} x_{t-1}
\end{flalign*}where $\epsilon_t\stackrel{iid}{\sim}\text{t}_{\beta_4}$, and $\text{t}_{\beta_4}$ denoting a standardized student-t random variable with $\beta_4$ degrees of freedom. We parameterize the error term $\epsilon_{t-1}$ in the volatility equation using the absolute value to cater for the heavy-tailed nature of the returns distribution. 

The above model yields an auxiliary likelihood for which the auxiliary scores (of the likelihood) can be easily calculated. As argued in \cite{martin2019auxiliary}, in state space models, such as the above $\alpha$-stable volatility model, the scores of auxiliary likelihoods yield convenient summary statistics for ABC-based inference, and in what follows we take as our choice of summary statistics for ABC the scores of this auxiliary GARCH model.

While the $\alpha$-stable volatility model has been used to fit returns data in several studies, it is one of a plethora of choices for modeling volatile return data. Herein, we apply the R-ABC approach, as well as the diagnostic devised in \cite{frazier2017model} to determine if the underlying $\alpha$-stable volatility model can adequately capture the features of daily returns series. 

\begin{figure}[h!]
	\centering
	\includegraphics[width=17cm, height=10cm]{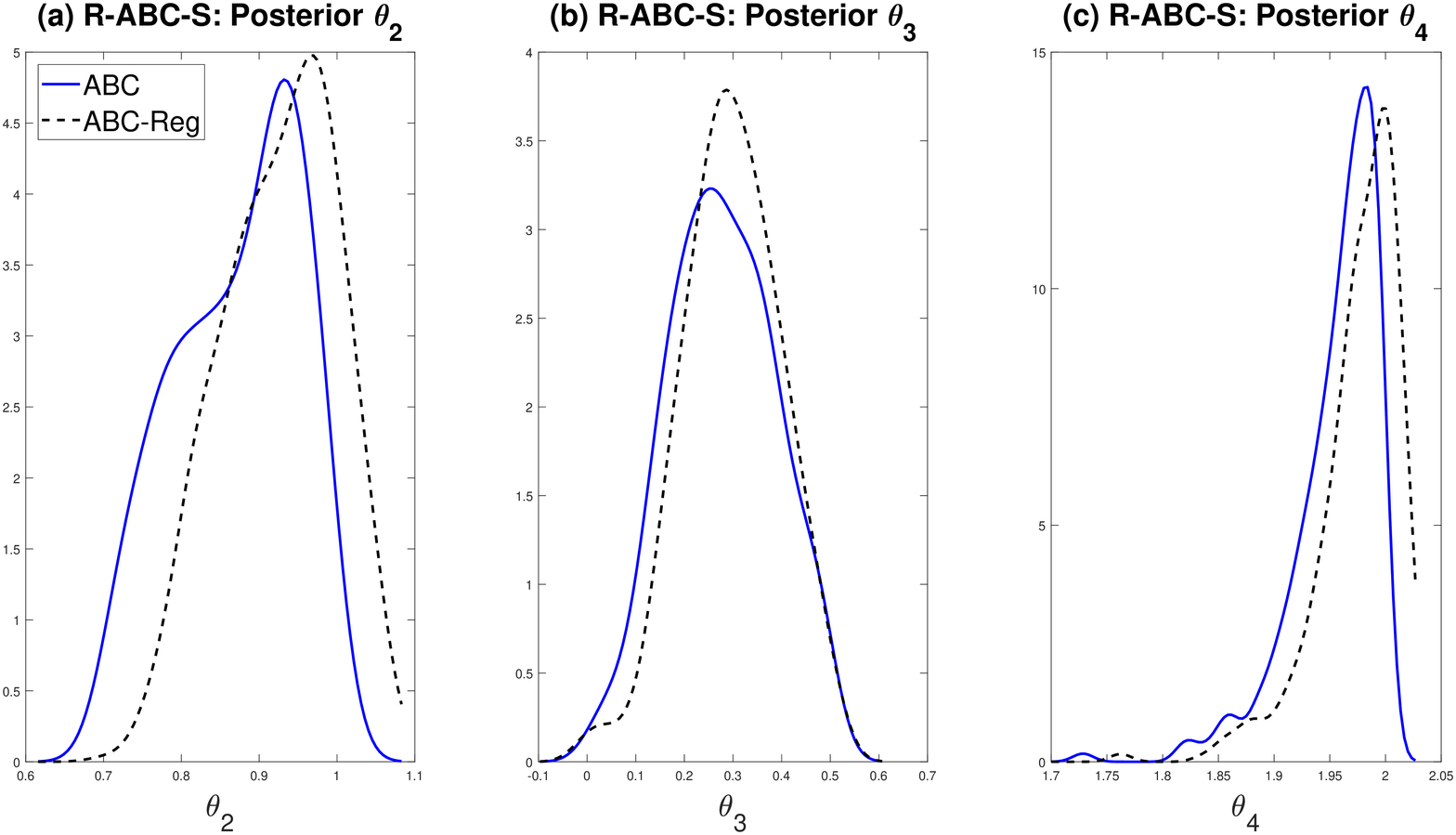} 
	\caption{Panels (a)-(c) provide the R-ABC-S posterior densities for $\theta_2,\theta_3,\theta_4$ using ABC (solid line) and ABC-Reg (dashed line). }%
	\label{Fig:alpha1}
\end{figure}

\begin{figure}[h!]
	\centering
	\includegraphics[width=17cm, height=10cm]{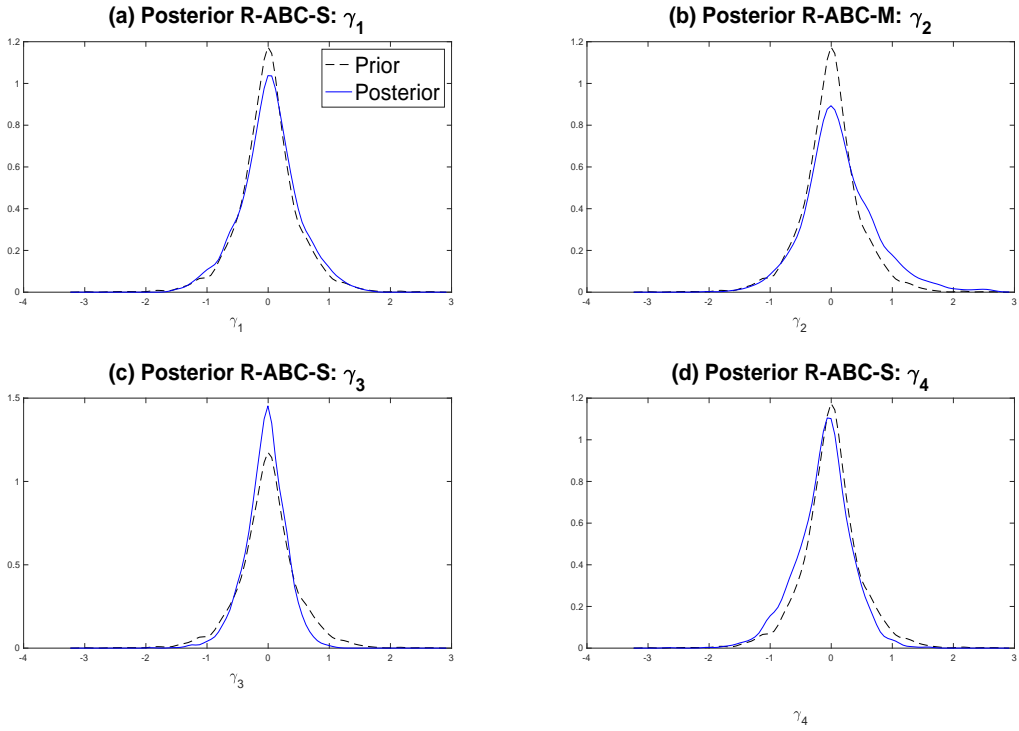} 
	\caption{Panels (a)-(d) provide the R-ABC-S posterior densities for the adjustment components $\gamma_1,\dots,\gamma_4$. The dotted line represents the prior, $\text{La}(0,\lambda=0.50)$, while the solid line is the posterior. }%
	\label{Fig:alpha2}
\end{figure}

We apply R-ABC-S using $N$=50,000 simulated data sets, with values in the smallest 1\% quantile of the simulated distances used to define the R-ABC posterior.\footnote{The results based on R-ABC-W are similar and are not reported for the sake of brevity.} For the adjustment parameters, $\Gamma=(\gamma_1,\dots,\gamma_4)'$, we again consider $\gamma_i\sim\text{La}(0,\lambda=0.50)$. We plot the R-ABC-S posteriors for the $\theta$ components in Figure \ref{Fig:alpha1}, which are very similar to those obtained by Marin et al. (2019). The posteriors for the adjustment components are given Figure \ref{Fig:alpha2} and do not significantly depart from the priors. 

These results suggest that the $\alpha$-stable volatility model captures the main features of the S\&P500 return series. To further demonstrate this fact, we compare the conclusions obtained from R-ABC with the graphical model misspecification diagnostic devised in \cite{frazier2017model}. This diagnostic graphically compares the behavior of the ABC acceptance probabilities against the relationship that should exist under correct model specification. Under correct model specification, \cite{frazier2016asymptotic} demonstrate that the relationship between the ABC tolerance and the ABC acceptance probability is roughly linear, however, when the model is misspecified this relationship becomes exponential; i.e., as the tolerance decreases the acceptance rate decreases at an exponential rate. 

\begin{figure}[h!]
	\centering
	\includegraphics[width=12cm, height=7cm]{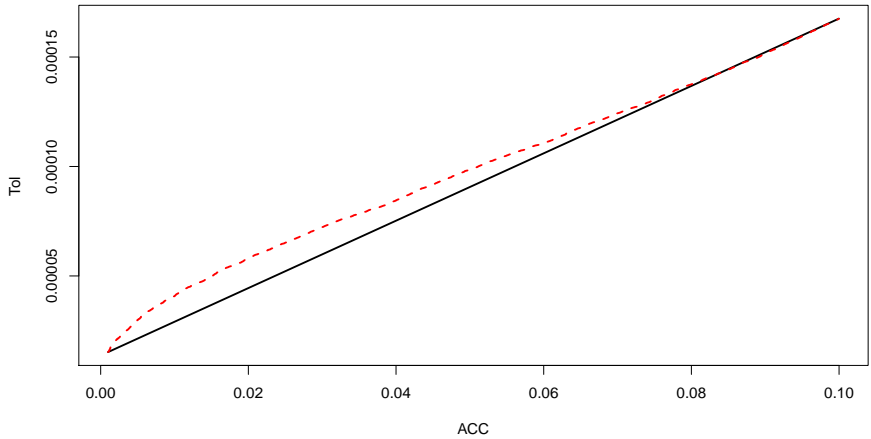} 
	\caption{Graphical diagnostic approach to detect model misspecification. The y-axis (Tol) corresponds to the tolerance used to select draws in ABC, and the x-axis (ACC) refers to the observed acceptance rate given the tolerance. Significant departures from the 45-degree line represents potential model misspecification.}%
	\label{Fig:alpha3}
\end{figure}

We plot the results of the graphical diagnostic in Figure \ref{Fig:alpha3}. This graphical diagnostic indicates that the resulting model is not significantly misspecified, and agrees with the conclusion obtained from analysing the posteriors for the R-ABC adjustment components.



\section{Discussion}
This paper has made two significant contributions to the literature on approximate Bayesian methods. Firstly, we have proposed an ABC approach that displays significant robustness to model misspecification, while being only slightly more computationally demanding than ABC. This new robust approach to ABC is based on augmenting either the summary statistics or the metric used in ABC with additional free parameters that can ``soak up'' any model misspecification.

Secondly, this new approach also allows the user to detect precisely which summary statistics are incompatible with the assumed data generating process. Incorporating this information within subsequent rounds of model building could lead to better models that can more accurately capture the behavior exhibited by the observed summary statistics. In this sense, the robust ABC approaches considered herein are similar to the model criticism approach of \cite{ratmann2009model}. In the context of ABC, \cite{ratmann2009model} propose an approach to detect aspects of the model that the summary statistics can not adequately capture. Their approach relies on treating the ABC tolerance as an unknown parameter, and augmenting the original ABC inference problem with this additional parameter. The authors argue that posterior realizations for the tolerance parameter that are ``large''  indicate the possibility of a mismatch between the model and the observed data, with ``large'' determined by some hypothesis test. 

While useful, in the case of multivariate summaries the approach of \cite{ratmann2009model} requires a tolerance parameter for each summary statistic used in the analysis, with posterior inference then required on the full set of model parameters and tolerance parameters, and subsequent corrections for multiple testing. As such, while the approach of \cite{ratmann2009model} is useful for model criticism, it is not clear how to easily benchmark the results to obtain a meaningful model misspecification indicator. In contrast, the approach considered herein has a direct benchmark with which to gauge the impact of misspecification on the summaries, namely the prior distribution of the adjustment components. If the corresponding posterior for the adjustment component does not resemble the prior, this is strong evidence that this summary can not be matched by the assumed model.

This new ABC approach is strongly motivated from the robust BSL approach proposed in \cite{FD2019}, where the authors use a similar idea to produce BSL posteriors that are robust to model misspecification. Given that BSL and ABC are both increasingly common tools in computational statistics, we believe these robust versions will be of great use to practicing statisticians.

\section*{Acknowledgements}
David T. Frazier gratefully acknowledges funding support by the Australian Research Council through grant DE200101070. Christopher Drovandi acknowledges funding support through Australian
Research Council Discovery Project DP200102101.

\bibliographystyle{apalike} 
\bibliography{refs}

\end{document}